\documentclass[12pt,a4paper]{article}

\usepackage{amsmath, amsthm, amssymb}
\usepackage{enumerate}
\usepackage{epsfig} 
\usepackage{graphicx}
\usepackage[T1]{fontenc} 
\usepackage[english]{babel}
\usepackage[applemac]{inputenc}
\usepackage{dsfont}

\newcommand{\IR}{\mathbb R}
\newcommand{\IP}{\mathbb P}

\newcommand{\sk}[1]{\left\langle #1\right\rangle}
\newcommand{\sn}[1]{\lvert\langle #1\rangle\rvert}

\newcommand{\sgn}{\mathrm{sgn}}

\theoremstyle{plain}

\theoremstyle{definition}

\begin{document}
\title{\vspace{-1.3cm}Could time-symmetric interactions reconcile relativity and quantum non-locality?}
\author{Dustin Lazarovici\thanks{Mathematisches Institut, Ludwig-Maximilians-Universit\"at, Therestr. 39, 80333 Munich, Germany. E-Mail: lazarovici@mathematik.uni-muenchen.de}}
\date{}
\maketitle
\begin{abstract}
\noindent We present a simple model demonstrating that time-symmetric relativistic interactions can account for correlations violating the Bell inequalities while avoiding conspiracies as well as the commitment to instantaneous influences. Based on an explicit statistical analysis of this model, we emphasize the essential virtues and problems of such an account and discuss its relation to Bell's  theorem. 
\end{abstract}
\section{Introduction}
\noindent In his beautiful article ``Speakable and unspeakable in quantum mechanics'' John S. Bell, discussing the implications of his seminal non-locality theorem \cite{Bell1, Bell2, Bell3} concludes:

\begin{quote}\textit{``For me then this is the real problem with quantum theory: the apparently essential conflict between any sharp formulation [of quantum theory] and fundamental relativity. That is to say, we have an apparent incompatibility, at the deepest level, between the two fundamental pillars of contemporary theory.... It may be that a real synthesis of quantum and relativity theories requires not just technical developments but radical conceptual renewal.''}~\cite{Bell5}\end{quote}

\noindent From todays perspective, there's only little to add to this assessment.  For one, we could emphasize that it's now well understood (and well established by various \textit{no-signaling} theorems for quantum mechanics) that quantum non-locality does not imply the possibility of faster-than-light \textit{communication} or any other way to violate the principles of relativity \textit{operationally}. One could also note that in the nearly two decades which have passed since Bell's statement, modest yet significant progress has been made towards generalizing sharp formulations of non-relativistic quantum mechanics to the relativistic regime. In fact, Bell himself in a later publication \cite{Bell4} suggested that the \textit{GRW collapse theory} may lend itself to a precise relativistic formulation, a feat that was indeed accomplished by Tumulka \cite{Tumulka}, albeit only in a non-interacting setting  (but see\cite{Bedingham} and references therein for further developments; See \cite{Esfeld} for a critical discussion of Tumulka's model).. Also, Lorentz invariant generalizations of \textit{Bohmian mechanics} can be formulated, although for the price of introducing a preferred foliation of space-time (which, however, can be generated by a Lorentz invariant law and shown to be empirically undetectable \cite{Durr}). All in all, the understanding that has grown over the last few years is that there is indeed not a \textit{contradiction} but a distinct \textit{tension} between relativity and quantum non-locality and that this tension is not primarily characterized by the thread of superluminal signaling (and the causal paradoxes that could result from it \cite{Bell3}) but, more subtly, by the fact that relativistic space-time -- having no structure of absolute simultaneity and no objective temporal order between space-like separated events -- is not particularly hospitable to the kind of \textit{instantaneous influences} that, according to Bell's theorem, the explanation of certain non-local correlations seems to require. (For a comprehensive discussion of the issue, see \cite{Maudlin}, \cite{Norsen}. See \cite{Gisin} for a simple argument substantiating said tension.)

 In this paper, we want to discuss the possibility to alleviate the tension between relativity and non-locality by explaining non-local correlations as a result of interactions that are both \textit{retarded}, ``propagating'' at a finite speed from past to future, and \textit{advanced}, ``propagating'' at a finite speed from future to past. Despite the technical difficulties inherent to such theories (the most important of which we will highlight in our discussion) and the philosophical reservations that one might have against retrocausality (but which shall not concern us here), the great virtue of such models would be that they could provide a complete physical account of quantum correlations while drawing exclusively on the resources of relativistic space-time with particles interacting along their \textit{past and future light-cones}. In fact, it can be argued on the basis of time-symmetry that advanced + retarded interactions are really the generic case of relativistic interactions and that it's the empirical \textit{violation} of this symmetry (e.g. the absence of advanced electromagnetic radiation) that requires explanation. And it has been speculated that such an explanation could parallel, or even reduce to, the statistical explanation of the thermodynamic asymmetry, accounting for the absence of advanced effects on \textit{macroscopic scales}, while the time-symmetry of the fundament laws may still reveal itself on \textit{microscopic} scales in the quantum phenomena \cite{Price}. (Indeed, such an analysis, trying to establish the empirical adequacy of a theory with advanced interactions, is not without precedence. In their ``absorber theory'', Wheeler and Feynman outline a statistical account of the radiative asymmetry based on a time-symmetric formulation of classical electrodynamics \cite{WF45, WF49}; See \cite{Dec1} for a recent, clarifying discussion.)

 The idea of accounting for quantum non-locality by admitting some form of retrocausality is not new, but has been advanced by various authors for many decades.\footnote{See, for instance, \cite{Beauregard, Beauregard2}, \cite{Stapp}, \cite{Davidon}, \cite{Sutherland}, \cite{Price}, \cite{Goldstein}. Some authors have proposed time-symmetric formulations of (non-relativistic) quantum mechanics, most notably Aharanov et.al. who developed a \textit{two-time formalism} \cite{Aharanov, Aharanov2} and J. G. Cramer, whose \textit{transactional interpretation} stipulates that any quantum mechanical interaction involves \textit{advanced and retarded} solutions of the wave-equation \cite{Cramer2, Cramer}. Concerning the latter, see \cite{Maudlin} for a critique and \cite{Lewis} for a recent review. Note also the time-symmetric formulation of Bohmian mechanics discussed by Sutherland \cite{Sutherland2}, and other references therein.} Nevertheless, it is rarely acknowledged as a possible implication of Bell's theorem and even rarer to pass the threshold from a logical possibility to a serious option. On the one hand, this is quite understandable, considering how drastically the proposal contradicts our ordinary sense of time and causation. On the other hand, it wouldn't be the first time that new physics require a radical revision of our respective (pre)conceptions. Hence, the best way to challenge the status quo is to develop and discuss further physical models that can serve as an intuition pump and demonstrate that retrocausation, in the appropriate context, must be neither as incomprehensible nor as unbecoming as generally assumed. This is precisely what the present paper aims to accomplish. 
 
By means of a simple toy-model, tailored, in particular, to the EPRB experiment, we want to demonstrate, in a quantitative manner, that time-symmetric relativistic interactions can {in principle} account for violations of the Bell inequalities \textit{without presupposing conspiratorial correlations} between the experimental parameters and the variables related to the preparation of the system. While retrocausal models are in general very difficult to analyze, we will present an explicit statistical treatment, showing how and in what sense relevant predictions can be extracted. 

Our toy-model, we emphasize, is neither able nor intended to reproduce all quantum spin statistics (for instance, it will fail to do so for repeated spin-measurements in varying directions), but is used to explore the range of logical possibilities that is restricted by Bell's theorem and to emphasize the chances and challenges associated with retrocausal accounts of non-locality. 

Finally, it should be needless to say, though we want to clarify nonetheless, that we do not engage in any futile attempt to evade the consequences of Bell's theorem. The question today is no longer \textit{if} the principle of local causality is violated in nature but \textit{how} -- and what this means for us as we move forward.

\newpage

\section{The model}
\noindent The model is defined by the following assumptions:

\begin{enumerate}[1)]
\item The particles have an internal degree of freedom (a ``hidden variable'') represented by a vector $\mathbf{S}$ on the 2-sphere $S^2$. We refer to this degree of freedom as the particle's \textit{spin}. 

\item A \emph{spin measurement} in the direction $\mathbf{a} \in S^2$ cannot determine the exact value of $\mathbf{S}$, but only its the orientation relative to $\mathbf{a}$. The result of a spin measurement is thus given by \begin{equation} \sgn \langle \mathbf{a} , \mathbf{S} \rangle \in \lbrace{\pm 1\rbrace},\end{equation} 
where $\sgn(x)$ denotes the sign of $x$. 

We say that the particle has \textit{a-spin up} if $\sk{\mathbf{a}, \mathbf{S}} >0$, i.e. $ \sgn \sk{\mathbf{a} , \mathbf{S}} = +1$ and  \textit{a-spin down} if $\sk{\mathbf{a}, \mathbf{S}} < 0$, i.e. $ \sgn \sk{\mathbf{a} , \mathbf{S}} = - 1$. 

We can disregard the special case $\sk{\mathbf{a}, \mathbf{S}}  = 0$, as it will have probability $0$. 

\item One of the crucial lessons of quantum mechanics is that a measurement is not a purely passive process, but an invasive interaction that will in general effect the state of the measured system. Here, in analogy to a projective measurement in quantum mechanics, we assume that a spin measurement influences the particle's hidden spin state by projecting it onto the respective direction in which it's being measured. That is, if the spin of the particle undergoing a spin measurement in the $\mathbf{a}$-direction is $\mathbf{S}$, its spin immediately after measurement will be
\begin{equation} \sgn \langle \mathbf{a} , \mathbf{S} \rangle \, \mathbf{a} =  \frac{\langle \mathbf{a} , \mathbf{S} \rangle}{\lvert \langle \mathbf{a} , \mathbf{S} \rangle \rvert} \, \mathbf{a}. \end{equation}

\item We are interested in the statistics of the EPRB-experiment concerning simultaneous measurements on a pair of entangled particles in the singlet state. 
To this end, we consider in our model an ensemble of pairs of particles whose initial spin-variables are prepared with opposite orientation in a random direction, which it equidistributed on the unit sphere $S^2$. For any pair, we denote
\begin{equation} \mathbf{S}^A(t=0) = - \mathbf{S}^B(t=0) = \mathbf{S}_0.\end{equation}
A spin-measurement usually ends with the particles being absorbed in a screen or a detector shortly after passing the magnet. 

\item The Spin state $\mathbf{S}$ is subject to a \textit{pair interaction} whose effect is such that a particle continuously rotates the spin of its partner towards the orientation antipodal to its own. This effect is manifested by an \textit{advanced and retarded} action of one particle on the other. This could mean e.g. that the interaction is transmitted by a medium -- like a field or a massless particle -- propagating with the speed of light towards \textit{past and future} or that the particles interact \textit{directly} along their \textit{past and future} light-cones. 

To acknowledge the fact that EPR correlations persist over very long distances \cite{Salart} -- indeed, if quantum mechanics is correct, over \textit{arbitrary} long distances -- we will have to assume that the interactions in a particle-pair are \textit{unattenuated}, i.e. unaffected by distance, and \textit{discriminating}, i.e. unfazed and unscreened by any other matter, thus realizing two essential properties of what Maudlin describes as the ``quantum connection'' \cite[p. 22]{Maudlin} or what, in other words, can be understood as a form of \textit{entanglement}. 

\end{enumerate}

\noindent To realize this model, we can for instance consider an interaction of the following type:\\

\noindent For two unit vectors $\mathbf{X}, \mathbf{Y}\in S^2\subset \IR^3$, their distance-vector is given by 
\begin{equation*} D(\mathbf{X},\mathbf{Y}) = \frac{\arccos \langle \mathbf{Y} , \mathbf{X} \rangle}{\sqrt{1- \langle \mathbf{Y} , \mathbf{X} \rangle^2}} \; \Bigl(\mathbf{Y}- \langle \mathbf{Y} , \mathbf{X} \rangle \mathbf{X}\Bigr) \in \mathrm{T}_\mathbf{X}S^2,\end{equation*}
where $\mathrm{T}_\mathbf{X}S^2$ denotes the tangent-bundle of the sphere at point $\mathbf{X}$. Then, assuming the particles have world-lines $z_1^\mu(t), z_2^\mu(t)$, we can set
\begin{equation}\label{ford} \frac{\mathrm{d}}{\mathrm{d}t} {\mathbf{S}}_i(t) \propto D\bigl(\mathbf{S}_i(t), - \mathbf{S}_j(\tau_{ret})\bigr) +  D\bigl(\mathbf{S}_i(t), - \mathbf{S}_j(\tau_{adv})\bigr), \end{equation} 

\noindent where $\tau_{ret}, \tau_{adv}$ are the \textit{advanced} and \textit{retarded} time, i.e. the solutions of 

\begin{equation}\label{taupm} \bigl(z_i^\mu(t) - z_j^\mu(\tau)\bigr)\bigl(z_{i, \mu} (t) - z_{j, \mu} (\tau) \bigr) = 0. \end{equation}

\noindent Hence, the spin of particle $i$ at time $t$ is ``repelled'' by the spin of particle $j$ at the advanced and retarded times. This is an example of a direct particle interaction along past and future light cones (fig. 1) (c.f. the Wheeler-Feynman theory of electrodynamics \cite{WF45, WF49}).

There is, however, one issue with this particular (most simple) example. Being defined by a first order differential equation, the law is actually not time-symmetric in the sense of \textit{time-reversal invariant}, thus undermining our best \textit{a priori} argument for admitting advanced interactions in the first place. 
In principle, this flaw could be resolved by considering the analogous second order equation, while adding appropriate constraints on the initial condition, namely that $\dot{\mathbf{S}}_1(0) = \dot{ \mathbf{S}}_2(0) = 0$ (on average). However, while qualitatively the same, the analysis of this interaction would be more complicated and our arguments will be considerably simplified by drawing on the first order differential equation. In general, however, there is a variety of conceivable interactions that could realize our model in the relevant respects, the most interesting of which might involve something like a quantum wave-function, though expanding on these possibilities would go beyond the scope of our discussion and, in fact, beyond the scope of this simple toy-model per se.

\begin{figure}[ht]
\begin{center}
    \includegraphics[scale=0.5]{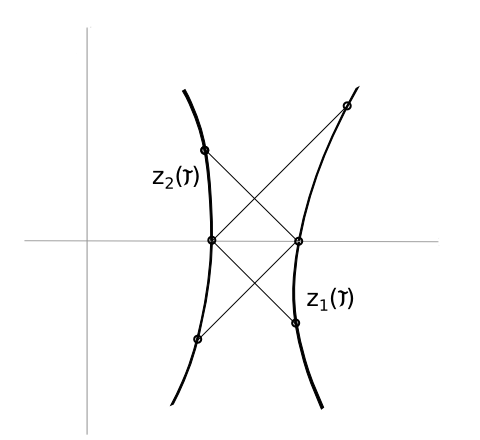}
\flushleft\caption{Direct interaction of two particles along past and future light-cones.}
\end{center}
\end{figure}


\section{A heuristic analysis of the time-evolution}

Figure 2 shows a sketch of the space-time diagram of the EPRB-experiment as described by our model. 
 \begin{figure}[ht]
 \begin{center}
 \includegraphics[width=13cm]{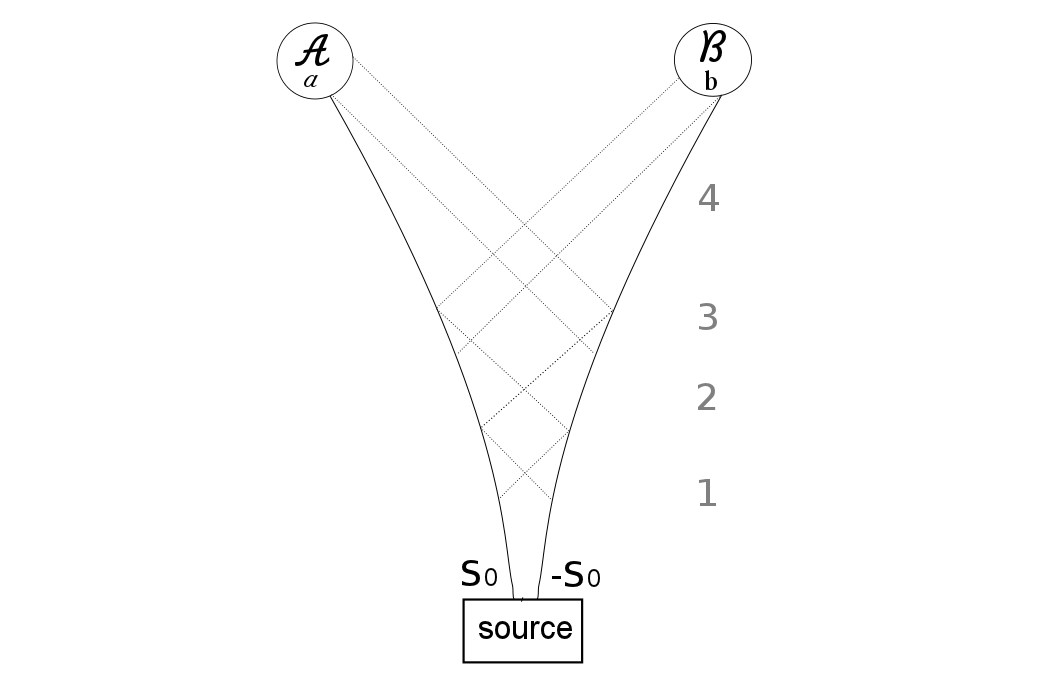}
\caption{Space-time diagram of the EPRB experiment. For the analysis of the interactions, we distinguish 4 increments of the particles world-lines in different slices of space-time.}
  \end{center}
 \end{figure}
\noindent A pair of particles prepared with opposite spin move in different directions towards an apparatus (a Stern-Gerlach magnet), where they undergo a spin measurement in directions $\mathbf{a}$ and $\mathbf{b}$, respectively. These parameters can be freely chosen by the experimentalists right before the measurement. Since the two measurements are assumed to happen simultaneously (in the laboratory frame), they occur in space-like separated regions of space-time, here denoted by $\mathcal{A}$ and $\mathcal{B}$. According to our model, the outcomes of these measurements are determined by the particles' hidden spin state as they pass the device. We denote the corresponding values by $\mathbf{S}_A$ and $\mathbf{S}_B$, respectively.  The result of the spin measurement on particle A in direction $\mathbf{a}$ is then given by $A:= \sgn \sk{\mathbf{a}, \mathbf{S}_A}$. The result of the spin measurement on particle B in direction $\mathbf{b}$ is $B:= \sgn \sk{\mathbf{b}, \mathbf{S}_B} $. 

In the end, we are interested in the probabilities of the \textit{coincidences} $A=B$, respectively the \textit{anti-coincidences} $A \neq B$. We recall that the Bell inequality (in its simplest version, assuming ``perfect'' anti-correlations for spin-measurements in the same direction as predicted by QM as well as our model) reads 
\begin{equation}\label{Bell} \IP(A \neq B \lvert a, b) + \IP(A \neq B \lvert b, c) + \IP(A \neq B \lvert a, c) \geq 1,\end{equation}
for arbitrary parameter-settings $a,b,c$ \cite{scholarpedia}. This inequality is violated for the correlations observed in the EPRB-experiment (for certain choices of $a,b,c$) thus implying, according to Bell's theorem, that they cannot be reproduced by any local theory.

Indeed, we can consider for starters the predictions of our hidden variable model \textit{without}  advanced interactions. This will serve as our point of reference as a completely \textit{local} model. Without advanced interactions, the initial configuration is a stationary point of the (retarded) dynamics and we have $\mathbf{S}_A = \mathbf{S}_0$ and $\mathbf{S}_B = -\mathbf{S}_0$, that is, the outcomes of the spin measurement depend only on the orientation of the initial hidden spin $\mathbf{S}_0$ relative to those of the measurement devices. Concretely, we find:
\begin{align*} A = B & \iff \sgn \sk{\mathbf{a}, \mathbf{S}_0} = \sgn \sk{\mathbf{b}, - \mathbf{S}_0}\\ 
A \neq B &\iff  \sgn \sk{\mathbf{a}, \mathbf{S}_0} = \sgn \sk{\mathbf{b}, + \mathbf{S}_0}.
\end{align*}
Of course, we now know from Bell's theorem that there can be no probability distribution of $\mathbf{S}_0$ for which such a model will reproduce statistical correlations violating the above inequality. And indeed, a short calculation shows that for the assumed equidistribution of $\mathbf{S}_0$ on $S^2$ and arbitrary (coplanar) angles with $a+b+c = 360^\circ$:
\begin{equation*} \IP(A \neq B \lvert a, b) + \IP(A \neq B \lvert b, c) + \IP(A \neq B \lvert a, c) = 1, \end{equation*}
\noindent and in general
\begin{equation*} \IP(A \neq B \lvert a, b) + \IP(A \neq B \lvert b, c) + \IP(A \neq B \lvert a, c) \geq 1, \end{equation*}
\noindent so that Bell's inequality is always satisfied. (More precisely, one can see by a few geometric considerations that $P(A \neq B \lvert a, b) = 1 - 2 \,\frac{\sphericalangle a, b }{\;360^\circ}$, where $\sphericalangle a,b \in [0, + 180^\circ]$ is the acute angle between the vectors $\mathbf{a}$ and $\mathbf{b}$.)

Let's now turn to the more interesting case and consider our model with retarded \textit{and} advanced interactions. The key difference, of course, is that now the particles \textit{post measurement}, whose states are affected by the external intervention of the measurement process, can have a retro-causal effect on the particles \textit{before} measurement. To extract statistical predictions for this case, we will have to understand how the final spin states $\mathbf{S}_A$ and $\mathbf{S}_B$, determining the experimental outcomes, result from the time-symmetric interactions between the particles. This will be essentially a combination of three different effects due to what is sometimes described as \textit{zigzag causality}:

 \begin{figure}[ht]
 \begin{center}
 \includegraphics[width=12cm]{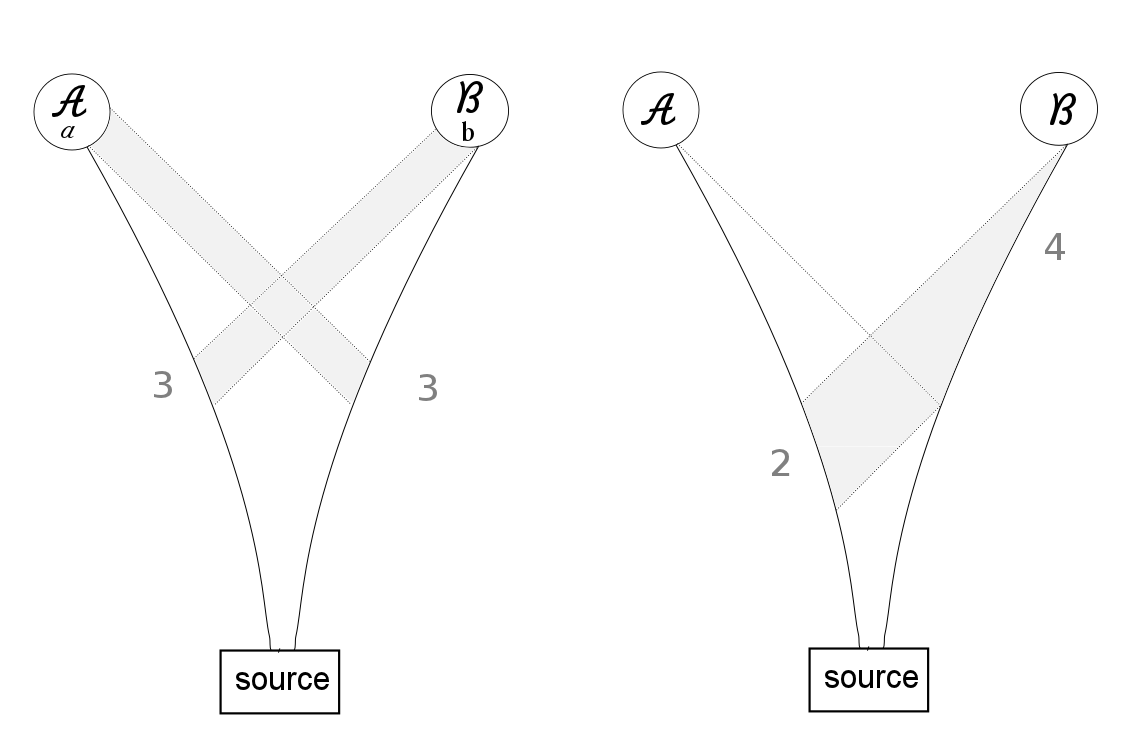}
\caption{Time-symmetric relativistic pair interactions in the EPRB setting. Left: ``feed-forward''. Right: ``preinforcement''.}
  \end{center}
 \end{figure}
 
\begin{enumerate}[1)]
\item \textit{Feed-forward:} Particle A after undergoing measurement in space-time region $\mathcal{A}$ exhibits an advanced action on its partner, reaching particle $B$ in space-time region 3 (fig. 3). The change in the spin-variable that particle $B$ experiences as a result is thus directed towards $-\mathbf{a}$ if the measurement on A yields \textit{a-spin up} and  $+\mathbf{a}$ if the measurement on A yields \textit{a-spin down}. 
The same holds vice-versa, with particle B in $\mathcal{B}$ exhibiting an advanced action on an earlier version of particle A, rotating its spin towards the direction opposite to B's post-measurement state. We call this effect a \textit{feed-forward} since, intuitively spoken, every particle receives a ``feedback'' of its partners future measurement result. 

\item \textit{Preinforcement:} After being affected by the \textit{feed-forward}, the advanced back-reaction of particle B in space-time region 4 on particle A in region 2 will tilt the spin of particle A slightly towards the direction \textit{in which it is going to be measured in the future.} The advanced action of particle A in space-time region $4$ has an analogous effect on particle B in space-time region 2. 
For further reference, we call this effect  \textit{preinforcement} -- so to speak, a preemptive reinforcement of the future measurement result. 

Of course, in a similar way, particle A in space-time region 2 has an advanced interaction with an earlier version of particle B, and so on... But further down the worldlines, the advanced effects originating from the particles after measurement are more and more diluted and will affect the state of the early particles only marginally.

\item \textit{Inertia:} Since the early spins are largely unaffected by the retro-causal influences, the effect of the \textit{retarded} action exhibited by the particles in space-time regions $1$ and $2$ is essentially to rotate the spins \textit{back} towards their initial values $\mathbf{S}_0$ and $-\mathbf{S}_0$, respectively.

\end{enumerate}

\newpage

\noindent In the course of these interactions, the hidden spin variable of particle A will vary within the convex set spanned by the vectors $\lbrace S_0, - B  \mathbf{b},   A \mathbf{a} \rbrace$, while the spin of particle B will lie within the antipodal triangle on $S^2$, spanned by $\lbrace - S_0, B  \mathbf{b},  - A  \mathbf{a} \rbrace$. (For the first order equation \eqref{ford}, these configurations form an invariant set, since $\dot{S}_i$ is always pointing into the interior of the respective triangle. For more general interactions, the following conclusion may hold only approximately or for short interaction times.) Hence, the spin configuration of particle A by the time it enters the measurement device will be of the form
\begin{align} 
\mathbf{S}_A =  \frac{\alpha\, \mathbf{S}_0  - \beta B  \, \mathbf{b} +  \gamma A \mathbf{a}}{\lVert\alpha\, \mathbf{S}_0  - \beta B  \, \mathbf{b} + \gamma\, A  \mathbf{a}\rVert},
\end{align}
with positive parameters $\alpha, \beta, \gamma; \, \alpha + \beta + \gamma = 1$, whose precise values in every run of the experiment will depend on the details of the interactions and the experimental setting. The parameters $\beta$ and $\gamma$ can thereby be thought of as reflecting the effect of \textit{feed-forward} and  \textit{preinforcement}, respectively. Since \textit{feed-forward} will, in general, act only for a brief period of time until the particles are absorbed in a detector and since \textit{preinforcement}, in turn, is only the ``echo'' of feed-forwards, we assume $\alpha > \beta > \gamma$. Analogously, $S_{B}$ will be of the form
\begin{align*}\mathbf{S}_B =  \frac{- \alpha'\, \mathbf{S}_0  - \beta' A \, \mathbf{a} + \gamma' B \mathbf{b}}{\lVert- \alpha'\, \mathbf{S}_0  - \beta' A \, \mathbf{a} + \gamma' B \mathbf{b}\rVert},
\end{align*}
and for simplicity we set $\alpha'=\alpha, \beta'=\beta$ and $\gamma'=\gamma$. 


For the results of the spin measurements, we have to consider the projections of the final spins $\mathbf{S}_A$ and $\mathbf{S}_B$ onto the corresponding directions in which they are being measured, that is (neglecting the normalization constant):
\begin{equation*}\begin{split}
\label{outcome}
\sk{\mathbf{a}, \mathbf{S}_A} &= \;\;\;\alpha \sk{\mathbf{a}, \mathbf{S}_0} -  \beta B \sk{\mathbf{a}, \mathbf{b}} + A \gamma \\
\sk{\mathbf{b}, \mathbf{S}_B}& =  - \alpha \sk{\mathbf{b}, \mathbf{S}_0} -  \beta A \sk{\mathbf{a}, \mathbf{b}} + B \gamma,
\end{split}
\end{equation*}
\noindent yielding
\begin{equation}\begin{split}
\label{outcomes}
A &=  \sgn\bigl\lbrace \;\;\; \alpha \sk{\mathbf{a}, \mathbf{S}_0} - B\, \beta \sk{\mathbf{a}, \mathbf{b}} + A \, \gamma\, \bigr\rbrace \\
B & =  \sgn\bigl\lbrace - \alpha \sk{\mathbf{b}, \mathbf{S}_0} - A\, \beta \sk{\mathbf{a}, \mathbf{b}} + B \gamma \;\bigr\rbrace.
\end{split}
\end{equation}

\noindent However, as we will see, this does not determine the outcomes unambiguously.

\section{Statistical Analysis}
\subsection{`Self-fulfilling prophecies' and the underdetermination of the time-evolution by Cauchy data}

The main technical and conceptual difficulty arising in theories with advanced and retarded interactions is that the laws of motion, in general, cannot be formulated as \textit{Cauchy problems}, meaning that the specification of \textit{initial conditions} at one single moment in time (respectively on a space-like hypersurface) is not sufficient to distinguish a unique solution and thus to determine the system's complete time-evolution. In our model, this problem is manifested in the fact that while for any choice of $\mathbf{a},\mathbf{b}$ there is a range of initial states $\mathbf{S}_0$ determining unique final states $\mathbf{S}_A$ and $\mathbf{S}_B$, there are also values of $\mathbf{S}_0$ (that is when $\sk{\mathbf{a}, \mathbf{S}_0}$ and/or $\sk{\mathbf{b},\mathbf{S}_0}$ are small compared to the effects of \textit{feed-forward} and \textit{preinforcement}) for which two or more \textit{different} prescriptions for $A$ and $B$ correspond to consistent evaluations of equation \eqref{outcomes} and thus to possible solutions of the equations of motion. Moreover, in these cases, there is a sense in which one could say that the measurement outcomes are \textit{retro-causally} responsible for their own occurrence. For further reference, we will call this phenomenon a \textit{self-fulfilling prophecy} (SFP).

We emphasize that this underdetermination of the time-evolution by initial data is not a result of our analysis being too coarse, but an intrinsic feature of theories admitting retro-causal influences.\footnote{For a mathematical discussion of this issue in the context of Wheeler-Feynman electromagnetism, see \cite{WF45, Dec1, Dec2} and references therein.} We also note that while the possibility of \textit{self-fulfilling prophecies} may be mind boggling, it need not imply the possibility of logical paradoxes  -- there is nothing inconsistent about the time-evolutions we consider here -- though the thread of potential inconsistencies is something to be addressed in the context of a more mature theory. Nevertheless, in addition to the philosophical headaches that might be caused by SFP, one very concrete difficulty that we have to face here is that theories in which solutions are not parametrized by Cauchy data are in general not \textit{statistically transparent}, in the sense that there is no obvious notion of a \textit{state-space} on which one could implement a \textit{statistical hypothesis} or define a measure of \textit{typicality}. More simply put, in our case, since the measurement outcomes $(A,B)$ are not unambiguously determined by the initial state $\pm \mathbf{S}_0$, their probabilities are not unambiguously determined by the statistical distribution of $\mathbf{S}_0$.

For this reason, we will have to resort to a more unconventional form of statistical analysis, leaving open the question what boundary conditions \textit{in addition} to $\mathbf{S}_0$ should be used to determine the time-evolution and how the resulting statistical description should look like in detail. While we cannot assign to each $(A,B) \in \lbrace \pm 1\rbrace^2$ a set of initial conditions $\mathbf{S}_0$ \textit{sufficient} to produce that outcome, thus implementing $(A,B)$ as a random variable on $S^2$, we will consider for each possible outcome $(A,B) \in \lbrace \pm 1\rbrace^2$ the \textit{necessary} conditions in terms of the initial configuration $\mathbf{S}_0$, thus determining \textit{upper} and \textit{lower bounds} on its probability. 


\subsection{A case-by-case analysis}

To this end, we will focus on the case $\sk{\mathbf{a},\mathbf{b}} < 0$, since quantum mechanics predicts that Bell's inequality \eqref{Bell} is always violated if the scalar product between any two of the (coplanar) vectors $\mathbf{a},\mathbf{b},\mathbf{c}$ is negative, with the most pronounced violation for $\sphericalangle a, b = \sphericalangle b,c, = \sphericalangle a,c = 120^\circ$.  Furthermore, observing that the problem is symmetric under $\mathbf{S}_0 \leftrightarrow - \mathbf{S}_0$ together with an exchange of the particle labels (and assuming that SFP respects this symmetry) we can w.l.o.g. assume $\sk{\mathbf{a}, \mathbf{S}_0} >0$. Now we can brake down all remaining cases corresponding to consistent solutions of eq. \eqref{outcomes} and thus to possible solutions of our assumed time-evolution. In the table below, we have listed for all $(A,B) \in \lbrace \pm1\rbrace^2$ the range of initial conditions $\mathbf{S}_0$ that \textit{can} produce the corresponding outcome:

 \begin{figure}[ht]
 \begin{center}
 \includegraphics[width=14cm]{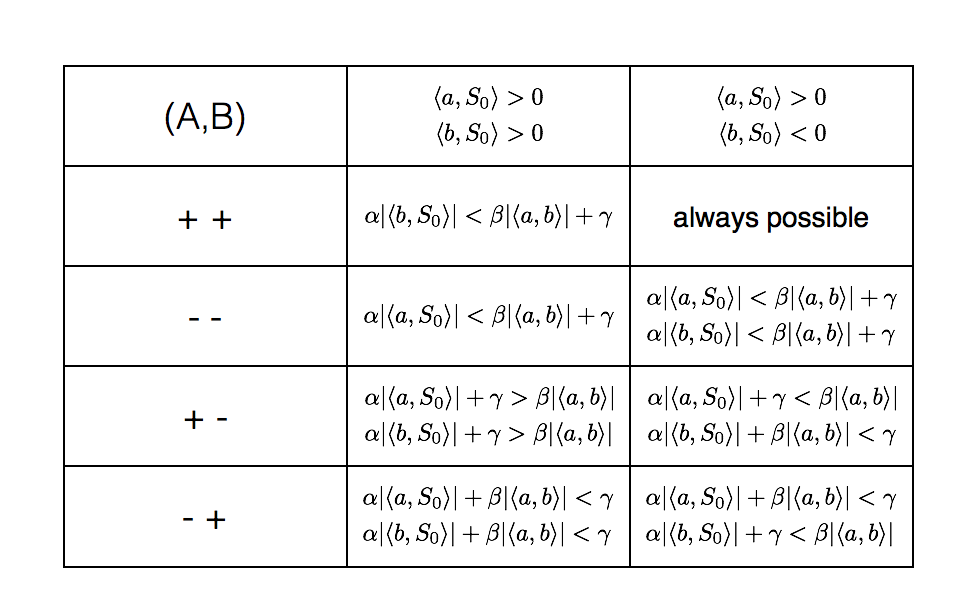}
 \vspace{-5mm}\caption{Table of possible outcomes for $\sk{\mathbf{a}, \mathbf{b}}<0$.}
 \end{center}
 \end{figure}

\noindent Admittedly, this may still seem quite confusing, but the dust will settle in a minute. First, we note that unless $\sk{\mathbf{a} , \mathbf{b}}$ is very close to $0$ or the ratio $\gamma/\beta$ unreasonably large, we will always find that $ \beta \sn{\mathbf{a}, \mathbf{b}}$ is greater or equal $\gamma$, meaning that we can disregard all the cases requiring $\beta \sn{\mathbf{a}, \mathbf{b}} < \gamma$. Furthermore, we recall that what we're ultimately interested in are the probabilities of the coincidences $A=B$ and the anti-coincidences $A\neq B$. To this end, only the following cases remain to be distinguished:\\

\noindent 1) If $\sgn\sk{\mathbf{a},\mathbf{S}_0}\neq\sgn\sk{\mathbf{b},\mathbf{S}_0}$ then $A=B$ occurs (almost surely).\\[1.5ex]
2) If $\sgn\sk{\mathbf{a},\mathbf{S}_0}=\sgn\sk{\mathbf{b},\mathbf{S}_0}$ then:
\begin{equation}\begin{split}\label{possib}
&A=B \text{ possible if } \Bigl( \alpha \sn{\mathbf{a},\mathbf{S}_0} < \beta \sn{\mathbf{a}, \mathbf{b}} + \gamma\, \vee\,\alpha \sn{\mathbf{b},\mathbf{S}_0} < \beta \sn{\mathbf{a}, \mathbf{b}} + \gamma \Bigr)\\
&A\neq B \text{ possible if } \Bigl(\alpha \sn{\mathbf{a},\mathbf{S}_0} + \gamma > \beta \sn{\mathbf{a}, \mathbf{b}} \, \wedge\, \alpha \sn{\mathbf{b},\mathbf{S}_0} + \gamma > \beta \sn{\mathbf{a}, \mathbf{b}} \Bigr).
\end{split}
\end{equation}

\noindent Note, in particular, that our model predicts perfect (anti-)correlations if the spins are measured in the exact opposite (respectively the same) direction.

 Comparing this to the local model without advanced interactions, where $A \neq B$ occurred if and only if $\sgn\sk{\mathbf{a},\mathbf{S}_0} = \sgn\sk{\mathbf{b},\mathbf{S}_0}$, we see that now, in certain cases that would have been ``on the edge'', i.e. where $\sk{\mathbf{a},\mathbf{S}_0}$ and  $\sk{\mathbf{b},\mathbf{S}_0}$ have the same sign but either one of the terms is small in absolute value, the spin variable of the particles are rotated just enough by the retro-causal \textit{feed-forward} to produce the coinciding event $A = B$ \textit{instead}.

Now a quantitative statistical analysis will, of course, require some information about the distribution of the parameters $\beta$ and $\gamma$. According to the specific form of the interactions, these parameters could vary in every run of the experiment, depending, in particular, on $\mathbf{S}_0$ and the control parameters $\mathbf{a}$ and $\mathbf{b}$. Going forward, we will make the simplest possible \textit{ansatz} which is that $\beta$ and $\gamma$ are not only the same for both particles in each pair, but constant throughout the ensemble of pairs, or, at least, distributed independently of the initial state $\mathbf{S}_0$ for fixed $\mathbf{a}$ and $\mathbf{b}$ (which is basically to say that the strength of the spin-interactions does not depend on the $S^2$-distance, i.e. the angle, between these spins). While one may object that this assumption is not only simple but somewhat simplistic, it will be fairly obvious that, qualitatively, our results will not depend on it too strongly. In any case, under this assumption, we can now determine an upper and lower bound on the probability of the anti-coincidences $A \neq B$ for given parameter choices $\mathbf{a}$ and $\mathbf{b}$ with $\sk{\mathbf{a}, \mathbf{b}}<0$. 

\begin{itemize}
\item The \textit{maximal probability} $\IP_{max}(A \neq B \lvert a, b)$ is the probability of $A \neq B$ assuming that $A\neq B$ will occur whenever it is \textit{possible} for an initial configuration $\mathbf{S}_0$ (or, in other words, that SFP  always favors $A\neq B$). 
\item The \textit{minimal probability} $\IP_{min}(A \neq B \lvert a,b)$ is the probability of $A \neq B$ assuming that it will occur only if $A = B$ is impossible for an initial configuration $\mathbf{S}_0$ (or, in other words, that SFP always favors $A=B$).  
\end{itemize}

\noindent From eq. \eqref{possib} we can conclude:
\begin{align*}
\IP_{max}(A \neq B \lvert a,b) &=  2\,\IP\Bigl( \sk{\mathbf{a},\mathbf{S}_0} > \frac{\beta \sn{\mathbf{a}, \mathbf{b}} - \gamma}{\alpha} , \sk{\mathbf{b},\mathbf{S}_0} > \frac{\beta \sn{\mathbf{a}, \mathbf{b}} - \gamma}{\alpha}  \Bigr)\\
\IP_{min}(A \neq B \lvert a,b) &=  2\,  \IP\Bigl( \sk{\mathbf{a},\mathbf{S}_0} > \frac{\beta \sn{\mathbf{a}, \mathbf{b}} + \gamma}{\alpha}, \sk{\mathbf{b},\mathbf{S}_0} >  \frac{\beta \sn{\mathbf{a}, \mathbf{b}} + \gamma}{\alpha} \Bigr),
\end{align*}

\noindent where the factor of $2$ accounts for the case $ \sk{\mathbf{a},\mathbf{S}_0}, \sk{\mathbf{b},\mathbf{S}_0} < 0$. This can be evaluated by means of the following identity that we derive in the appendix:
\begin{equation}\label{toprove} \bigl\lvert S^2 \bigr\rvert^{-1} \, \bigl\lvert \bigl\lbrace S \in S^2 : \sk{\mathbf{a},\mathbf{S}} > C \wedge \sk{\mathbf{b}, \mathbf{S}} > C\bigr\rbrace\bigr\rvert = \frac{1}{\pi} \int\limits_C^{\sqrt{\frac{1}{2}(1+\sk{\mathbf{a},\mathbf{b}})}} \sqrt{\frac{z^2-C^2}{z^2-z^4}}\, \mathrm{d}z. \end{equation}

\noindent For better illustration of the results, we will simplify things a bit further, still, by estimating the ratio of $\beta$ to $\gamma$, thus obtaining a parameterization of the final spin states $\mathbf{S}_A$ and $\mathbf{S}_B$ in terms of a single affine parameter $\nu \in [0,1]$. This is to say that we differentiate the analysis only by the \textit{strength} of the advanced effects -- represented by the parameter $\nu$ -- rather than by their dependence on distance or the duration of action, which could be dissected in a more fine-grained analysis by fitting the parameters $\beta$ and $\gamma$ (cf. figure 7 below). Setting $\beta = \nu$, a reasonable estimate is $\gamma = \nu^2$ since \textit{preinforcement} is a result of \textit{two} advanced interactions, and hence $\alpha = 1-(\beta + \gamma) = (1 - \nu - \nu^2)$.

\subsection{Results}
In figure 6, we have plotted the minimum and maximum probabilities for varying values of $\nu$ and parameter settings $\sphericalangle a,b = 120^\circ$, i.e. $\sk{\mathbf{a}, \mathbf{b}} =- \frac{1}{2}$, which is the case for which quantum mechanics predicts the greatest violation of the Bell inequality \eqref{Bell}. The upper curve represents the highest possible probability, the lower curve the lowest possible probability of the anticoincidence $A \neq B$, depending on the value of $\nu$, and we see that both are \textit{decreasing} with $\nu$, i.e. as retro-causal effects get more pronounced. The shaded area in between thus corresponds to the range of possible probabilities that could result from the present model, depending on how SFP's are fixed. The growing spread between $\IP_{min}$ and $\IP_{max}$ is due to the fact that with increasing $\nu$, the effects of \textit{preinforcement} become more and more relevant and \textit{self-fulfilling prophecies} thus more and more likely.

\begin{figure}[ht]
\begin{center} 
 \includegraphics[width=11cm]{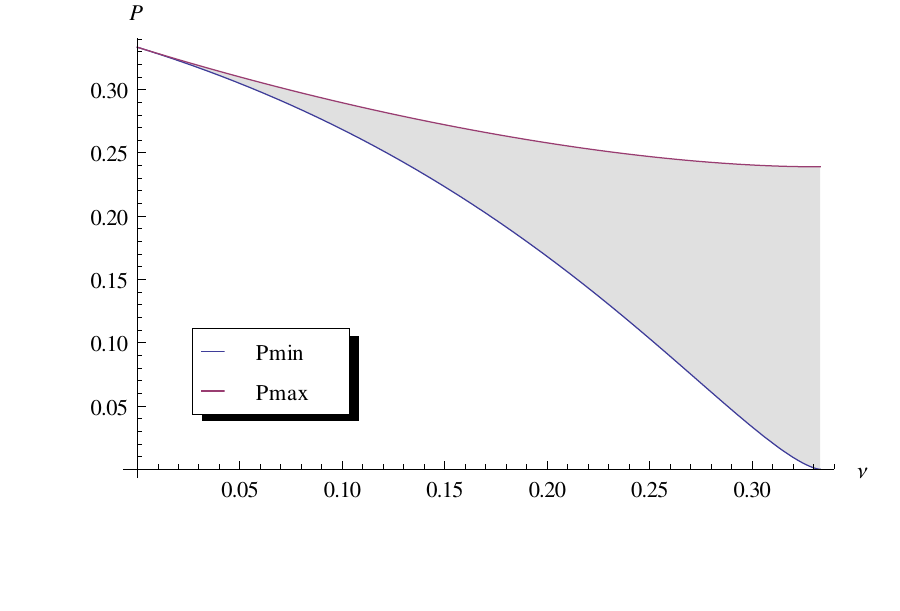}
 \label{Probminmax}
\vspace{-1cm} \caption{Possible values for $\IP(A\neq B \lvert a = 0^\circ, b = 120^\circ)$. }
\end{center}
\end{figure}

We see that for $\nu \equiv 0$, i.e. in the \textit{absence} of advanced interactions, the account reduces to the local model discussed in section 3. That is, $\IP_{min}$ and $\IP_{max}$ coincide (since $(A,B)$ now \textit{is} a random variable on $S^2$) yielding, in particular, $\IP(A \neq B \lvert \sphericalangle a, b = 120^\circ) = \frac{1}{3}$ and thus for \mbox{$a = 0^\circ, b=120^\circ, c = 240^\circ$}:
\begin{equation*} \IP(A \neq B \lvert a, b) + \IP(A \neq B \lvert b, c) + \IP(A \neq B \lvert a, c) = \frac{1}{3} + \frac{1}{3} + \frac{1}{3} = 1. \end{equation*}

\noindent Hence, the Bell inequality is, of course, satisfied, though it's important to note that this case is already critical, i.e. that equality holds in eq. \eqref{Bell}. For now, as we consider values of $\nu$ \textit{greater} $0$, we see that the statistical effect of the advanced interactions is to \textit{lower} the probability of the anticoincidence $A \neq B$, thus leading to a \textit{violation} of the Bell inequality. Notabene, the fact that the maximal probability $\IP_{max}$ already is smaller than $\frac{1}{3}$ shows that the violation of the Bell inequality is not \textit{because of} self-fulfilling prophecies, i.e. is not achieved by exploiting the underdetermination of the outcome by the initial state, but would have to occur \textit{in any case}, regardless of how this underdetermination is resolved and what the resulting statistical description would like in detail. 

On a more quantitative note, we have to keep in mind that we've made a series of simplifications and assumptions along the way, so it may or may not be significant that we find the range of possible probabilities to be fairly close to the quantum mechanical prediction of $0.25$ for reasonable values of~ $\nu$. 

To settle on more precise predictions, we can consider the \textit{median} of $\IP_{min}(A\neq B)$ and $\IP_{max}(A\neq B)$, corresponding to the probability of the anti-coincidence assuming that SFP is not biased between $A=B$ and $A\neq B$. These values are plotted in Figure 7 -- now, for greater generality, with the parameters $\beta$ and $\gamma$ varying independently -- to be compared with the predictions of the local model ($P=1/3$) and of quantum mechanics ($P= 1/4$). One of the most interesting things to note about this plot is that it doesn't look particularly interesting -- the values are rather steady except for the upward slope as $\gamma$ and $\beta$ go to $0$ and the downward slope as they become very large -- indicating a remarkable robustness of the predictions of the time-symmetric model against the details of the interactions.
\vspace*{-4mm}
\begin{figure}[ht]
 \begin{center}
 \includegraphics[width=10cm]{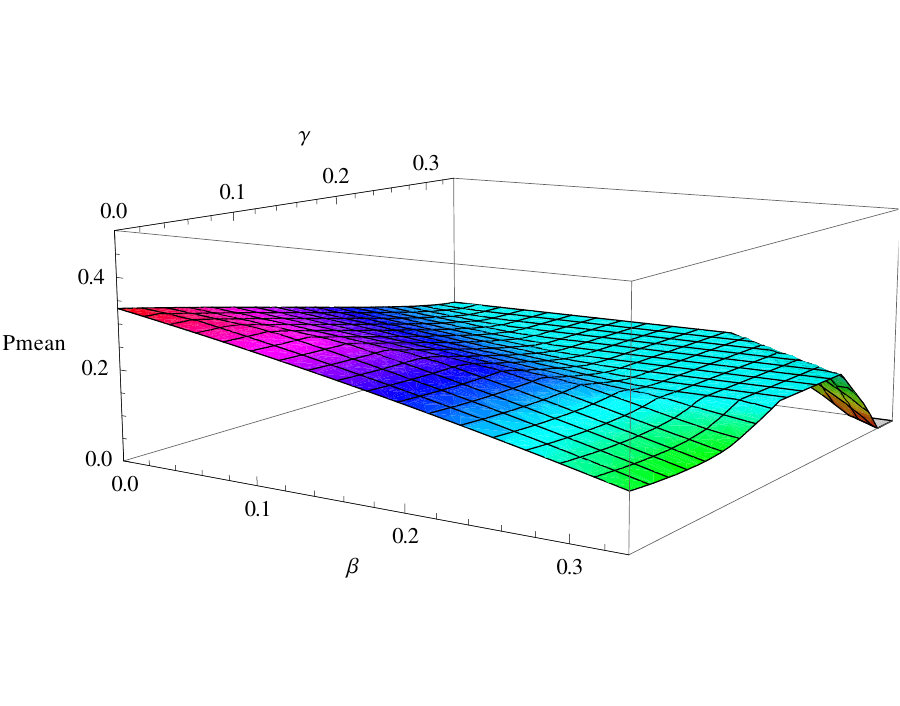}
 \label{Probminmax}
 \vspace*{-5mm}\caption{Mean Probability for different values of $\beta$ and $\gamma$.}
 \end{center}
\end{figure}

\noindent Finally, we note that there is no \textit{fixed} choice of parameters $\beta$ and $\gamma$ for which this particular evaluation reproduces the predictions of quantum mechanics (for the spin-singlet state) for \textit{all} values of $a$ and $b$, that is
\begin{equation}\label{quantumpred} \IP(A \neq B \mid a,b) = \frac{1}{2} + \frac{1}{2}\sk{\mathbf{a}, \mathbf{b}}. \end{equation} 
Of course, it would've been quite miraculous if it \textit{did}, given the overall crudeness of our statistical hypothesis. An interesting question might thus be whether by taking into consideration the possible dependence of $\beta$ and $\gamma$ on the relevant physical variables, one could find a probability distribution $p(\beta, \gamma \lvert \mathbf{S}_0,  \mathbf{a}, \mathbf{b})$ for which the model reproduces eq. \eqref{quantumpred} exactly.\\

In conclusion, what our analysis shows is that the quantum non-locality we observe in nature in form of statistical correlations violating Bell's inequality could really be understood as the signature of retro-causal effects due to time-symmetric relativistic interactions, rather than instantaneous (or superluminal) influences between space-like separated events.


\section{Retro-causality and Bell's Theorem}
Having seen that time-symmetric relativistic interactions can, in principle, account for the violation of the Bell inequality, it is very instructive to reflect on how exactly such a model would fit into the framework of Bell's non-locality theorem. We recall that the most general derivation of a Bell inequality (more specifically the CHSH inequality \cite{Bell2, CHSH}) is based on two (and only two) assumptions:

\begin{enumerate}[i)]
\item \textit{The locality assumption}:  The statistical correlations 

\begin{equation}\label{correlations} \IP(A , B \lvert a,b) \neq \IP(A \lvert a) \IP(B \lvert b) \end{equation} 

between the outcomes of the space-like separated measurement events in the EPR experiment are \textit{locally explainable}. By Bell's definition, a candidate theory provides a \textit{local explanation} of the correlations if conditioning on all the physical data in the (causal) past of $A$ and $B$ which, according to that candidate theory, could be relevant to the prediction of $A$ and $B$, will \textit{screen off} the correlations \eqref{correlations}, meaning that the specification of $A$ and $a$ becomes redundant for the prediction of the probability of $B$, and vice versa. Formally, comprising all possible ``common causes'' of $A$ and $B$ in a set of variables $\lambda$, the \textit{locality condition} reads:
\begin{equation}\begin{split}\label{locality}
&\IP(A \lvert B , a, b, \lambda) = \IP(A \lvert a, \lambda)\\ 
&\IP(B \lvert A, a, b, \lambda) = \IP (B \lvert b, \lambda).
\end{split}\end{equation}


\item {The no-conspiracy assumption}: The explanation of the correlations must not be \textit{conspiratorial}, meaning that the experimental parameters $a$ and $b$ can be chosen \textit{freely} or \textit{randomly}, independent of each other and of any other physical process that might be relevant to the system before measurement, hence independent of $\lambda$. Formally:
\begin{equation}\label{noconsp}
\IP(\lambda \lvert a, b) = \IP(\lambda).
\end{equation}
\end{enumerate}
 \begin{figure}[ht]
 \begin{center}
 \includegraphics[width=10cm]{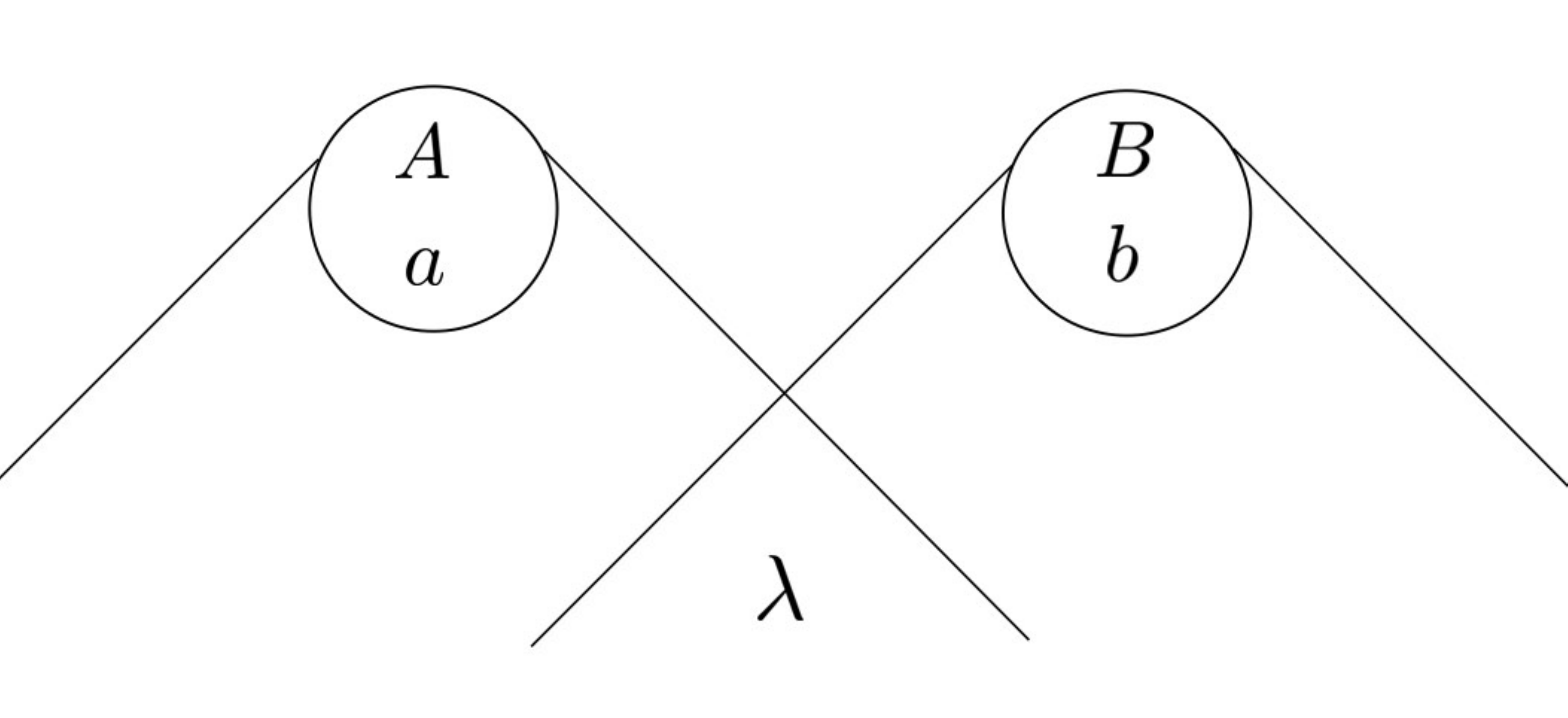}
\caption{Space-time diagram of a local explanation of EPR correlations.}
 \end{center}
 \end{figure}

\noindent Since our model violates the Bell inequality (as well as the CHSH inequality) it must violate at least one of these assumptions. In some discussions, a ``retrocausal explanation'' of the EPR correlations is  understood virtually synonymous with a ``conspiracy''. However, the description provided by our toy-model is, all things considered, reasonable enough to show that the issue deserves a second look and indeed, on that second look things turn out to be a bit more subtle: 

If, in our model, we condition the measurement outcomes on the relevant physical configurations in the \textit{past} of  $\mathcal{A}$ and $\mathcal{B}$ -- that is, on the particles' initial spin-variables $\pm \mathbf{S}_0$ -- we find that the locality condition \eqref{locality} (in form of ``parameter-independence'' \cite{Jarrett}) is violated, i.e.
\begin{align*} \IP(A \lvert a,b, \mathbf{S}_0) \neq \IP(A \lvert a, \mathbf{S}_0)\\
\IP(B \lvert a,b, \mathbf{S}_0) \neq \IP(B \lvert b, \mathbf{S}_0)
\end{align*}

\noindent This initial data, in other words, is not sufficient to screen off the correlations between the measurement outcomes on one side of the experiment and the parameter settings on the other. Hence, a physicist seeking to explain the correlations between the distant events, as we usually do, by looking for a common cause in their past is bound to fail, and may reasonably conclude that there must be some sort of instantaneous influence between the two sides of the experiment. However, if we lived in a world guided, on the microscopic level, by time-symmetric relativistic laws, this physicist, in doing so, would literally miss half the story, since the physical laws were such that the outcomes of the spin measurements were actually determined by the physical configurations in both past and future of $A$ and $B$. 

On the other hand, if we condition $(A, B)$ on the initial states $\pm \mathbf{S}_0$ \textit{and} the particle states \textit{after} measurement -- which, in a time-symmetric model like ours, can actually be regarded as ``causing'' the outcomes --  the probability, trivially, factorizes and the locality condition is, formally, satisfied. 

Note that no assumption about the localization of $\lambda$ actually enters the derivation of the CHSH inequality, though it is only in the case that $\lambda$ refers to configurations or events in the causal \textit{past} of $\mathcal{A}$ and $\mathcal{B}$ that we would speak of a local explanation in Bell's sense. This is to say, in particular, that the concept of locality or local causality presupposes a distinguished direction of time. From a time-symmetric perspective, a more natural desideratum (at least for theories that are not intrinsically stochastic) would simply be the \textit{absence of direct influences between space-like separated events.} And while our model is explicitly non-local in the conventional sense, it does satisfy the latter requirement which makes it arguably as relativistic as a non-local theory can get. 

In any case, from \textit{this} point of view, our model violates the Bell inequalities -- as it cannot be otherwise -- by violating the no-conspiracy condition. Obviously, the physical variables screening off the correlations are not independent of the parameter settings, since they include the particle states \textit{post} measurement which are collinear with the chosen orientations $\mathbf{a}$ and $\mathbf{b}$ of the measurement devices. However, we see no reason to deem such an account ``conspiratorial'' -- at least not in the devastating sense argued by Bell \cite{Bell2, Bell3} and others \cite{CHSH2} to essentially render futile the scientific enterprise. The fact that the state of the microscopic system after we have interacted with it reflects our experimental choices is hardly mysterious and in no way different from what we have anyway come to expect. More importantly, the advanced effects of the microscopic interactions do in no way infringe upon our freedom to choose the parameters of the experiment as we please (or make the choice completely random), nor on the possibility to prepare a system (it's initial state, that is) according to a our liking and practical abilities. In other words, while formally violating the no-conspiracy condition, the account does not presuppose any dependence between the parameters associated with the preparation of the system and the parameters associated with the setup of the measuring apparatus. Hence, it involves no conspiracy. \\

\noindent \textbf{Acknowledgements}: I am grateful to Christian Beck as well as anonymous referees for insightful comments on earlier versions of this manuscript.



\newpage

\begin{appendix}
\section*{Appendix: Derivation of equation \eqref{toprove}}
We want to compute the measure of the set $\bigl\lbrace \mathbf{S} \in S^2 \mid \sk{\mathbf{a},\mathbf{S}} > C, \sk{\mathbf{b}, \mathbf{S}} > C\bigr\rbrace$, for $C>0$ and $\sk{\mathbf{a}, \mathbf{b}}<0$.
In spherical coordinates, the variable $\mathbf{S} \in S^2 \subset \IR^3$ is parameterized as 
\begin{equation*} \mathbf{S} = ( \sin \theta \cos \phi, \sin \theta \sin \phi, \cos \theta), \; \theta \in [0, \pi), \phi \in [0, 2 \pi). \end{equation*}
W. l.o.g. we can locate $\mathbf{a}$ and $\mathbf{b}$ in the $x$-$y$-plane and set 
\begin{equation*} \mathbf{a} = (0 , 1 , 0);\, \mathbf{b} = (\sin\chi, \cos\chi, 0), \end{equation*} 
where $\chi$ is the angle between $\mathbf{a}$ and $\mathbf{b}$ and hence $\sk{\mathbf{a},\mathbf{b}}= \cos \chi$. We thus have
\begin{align*}
&\sk{\mathbf{a}, \mathbf{S}} = \sin \theta \sin \phi\, {>} C \\
&\sk{\mathbf{b}, \mathbf{S}} = \sin \theta \bigl( \sin \chi \cos \phi + \cos \chi \sin \phi)\, {>} C.
\end{align*}
Since the set is symmetric under interchange of $\mathbf{a}$ and $\mathbf{b}$, it suffices to consider the case $ \sk{\mathbf{b}, \mathbf{S}} > \sk{\mathbf{a}, \mathbf{S}} $ (and then double the measure).\\ 
Since the set is symmetric under reflection on the $x$-$y$-plane, it suffices to consider the case $\theta < \frac{\pi}{2}$, i.e. $S_z >0$ (and then double the measure). 

\noindent Then we have, for once,
\begin{align*}0< \sk{\mathbf{a}, \mathbf{S}} < \sk{\mathbf{b}, \mathbf{S}} & \iff 0 < \sin\phi <  \sin \chi \cos \phi + \cos \chi \sin \phi\end{align*}
which yields $ \cos\phi > 0$ and, after a little bit of algebra,
\begin{equation*}0 < \sin \phi < \sqrt{\frac{1}{2} \bigl(1 + \cos \chi\bigr)} = \sqrt{\frac{1}{2} \bigl(1 + \sk{\mathbf{a}, \mathbf{b}}\bigr)} =:D. \end{equation*} 
And we compute for $M:= \bigl\lbrace \mathbf{S} \in S^2 \mid  \sk{\mathbf{b}, \mathbf{S}} > \sk{\mathbf{a},\mathbf{S}} > C, \, S_z >0 \bigr\rbrace \subset S^2$:
\begin{align*} \bigl \lvert M \bigr\rvert &= \int\limits_{0}^{\pi/2} \int\limits_{0}^{2\pi} \mathds{1}_M(\theta, \phi) \sin \theta\, \mathrm{d}\theta\, \mathrm{d}\phi =  \Bigl\lvert \int\limits_{0}^{2\pi}  \int\limits_{0}^{1} \mathds{1}\lbrace  \sin\theta > \frac{C}{\sin\phi},\,  \sin\phi < D, \phi \in (0, {\pi}/{2}) \rbrace\, \mathrm{dcos} \theta \,\mathrm{d}\phi \, \Bigr\rvert\\
&= \int\limits_{0}^{\pi/2} \mathds{1}\lbrace C<\sin\phi<D \rbrace\, \sqrt{1-\frac{C^2}{\sin^2\phi}}\;\;  \mathrm{d}\phi = \int\limits_{C}^D \sqrt{\frac{z^2 - C^2}{z^2-z^4}}\; \mathrm{d}z,
\end{align*}
where in the last step we substituted $z:= \sin\phi$. Together with $\lvert S^2\rvert = 4 \pi$, equation \eqref{toprove} follows.

\end{appendix}


\newpage


\begin{thebibliography}{11pt}
    
   
     \bibitem{Aharanov} Y. Aharonov and B. Reznik, \grq On a Time Symmetric Formulation of Quantum Mechanics'. \textit{Phys. Rev. A} 52, 2538, (1995).
    
    \bibitem{Aharanov2} Y. Aharonov and E.Y. Gruss  \grq Two-time interpretation of quantum mechanics'. Preprint: \textit{arxiv: 0507269}  [quant-ph] (2005).
       
    	  \bibitem{Bedingham} Bedingham, D., D\"urr, D., Ghirardi, G., Goldstein, S., Tumulka, R. \& Zangh\'i, N. 2014 Matter Density and Relativistic Models of Wave Function Collapse. \textit{J. Stat. Phys.} \textbf{154}, 623--63. (doi: 10.1007/s10955-013-0814-9)
    
       \bibitem{Bell1} J.S. Bell, \grq On the Einstein-Podolsky-Rosen paradox', \textit{Physics} \textbf{1} (1964); reprinted in \cite[pp. 14--21]{Bell}
      
       \bibitem{Bell6} J.S. Bell,  \grq The theory of local beables'. \textit{Epistemological Letters} (1976); reprinted in \cite[pp. 52--62]{Bell}.
      
      
	\bibitem{Bell2}{J.S. Bell,  \grq Bertlmann's socks and the nature of reality'. \textit{Journal de Physique}, Colloque C2, suppl. au numero 3, Tome 42 (1981); reprinted in \cite[pp. 139 -- 158]{Bell}.}
	
  \bibitem{Bell5} J.S. Bell, \grq Introductory Remarks'. \textit{Physics Reports} 137, 7, (1986); reprinted in \cite[pp. 169--172]{Bell}.
	
 
\bibitem{Bell4} J.S.  Bell, \grq Are there quantum jumps?'. in: C. W. Kilmister (ed.), \textit{Schr\"odinger: Centenary of a polymath}, Cambridge  (1987); reprinted in \cite[pp. 201--212]{Bell}.
 
 
\bibitem{Bell3}{J.S. Bell,  \grq La nouvelle cuisine', in: A. Sarlemijn and P. Kroes (eds.), \textit{Between Science and Technology}, Elsevier Science Publishers (1990);  reprinted in \cite[pp. 232 -- 248]{Bell}.}
	
	
\bibitem{Bell} J.S. Bell,  \textit{Speakable and Unspeakable in Quantum Mechanics}. 2nd Edition. Cambridge Univ. Press, Cambridge (2004).

\bibitem{Beauregard} O. Costa de Beauregard, \grq M\'echanique quantique'. \textit{Comptes Rendus Acad\'emie des Sciences} 236, 1632, (1953).

    \bibitem{Beauregard2}O. Costa de Beauregard, \grq Time symmetry and the Einstein paradox.' \textit{Il Nuovo
Cimento} 42B, 41--64, (1977).
    
    

    \bibitem{Cramer2}J.G. Cramer,  \grq Generalized absorber theory and the Einstein-Podolsky-Rosen paradox'. \textit{Phys. Rev. D} 22 (2), pp. 362--376, (1980). 
    
    \bibitem{Cramer}J.G. Cramer,  \grq The transactional interpretation of quantum mechanics'. \textit{Rev. Mod. Phys.} 58 (3), 647--687, (1986).
    
    \bibitem{Davidon}W.C. Davidon, \grq Quantum physics of single systems'. \textit{Il Nuovo Cimento} 36B, 34--39, (1976).
    
    \bibitem{Dec1}{D.-A. Deckert, \textit{Electrodynamic Absorber Theory}. Der Andere Verlag. ISBN 978-3-86247-004-4, T\"onning (2010).}
    
    \bibitem{Dec2}D.-A. Deckert, D. D\"urr and N. Vona, \grq Delay Equations of the Wheeler-Feynman Type'. \textit{arXiv:1212.6285} [math-ph] (2012).
    
    \bibitem{Durr}D. D\"urr, S. Goldstein, T. Norsen, W. Struyve and N. Zangh\'i,  \grq Can Bohmian mechanics be made relativistic?'. Preprint: \textit{arXiv:1307.1714} [quant-ph] (2013). 
    
    
    \bibitem{EPR} A. Einstein, B. Podolsky  and N. Rosen, \grq Can quantum-mechanical description of reality be considered complete?' \textit{Phys. Rev.} 47, 777--780, (1935)
    
    \bibitem{Esfeld} M. Esfeld and N. Gisin, \grq The GRW flash theory: a relativistic quantum ontology of matter in space-time?'. Preprint: \textit{arXiv:1310.5308} [quant-ph] (2013).
    
        \bibitem{Goldstein}  S. Goldstein and R. Tumulka, \grq Opposite arrows of time can reconcile relativity and nonlocality'. \textit{Quantum Grav.} 20, 557, (2003).
    
   \bibitem{Gisin} Gisin, N. \grq Impossibility of covariant deterministic nonlocal hidden-variable extensions of quantum theory', \textit{Phys. Rev. A} 83, 020102(R) (2011).
    
    
   \bibitem{scholarpedia} S. Goldstein et.al.  \grq Bell's theorem', \textit{Scholarpedia, 6(10):8378}, http://www.scholarpedia.org/article/Bell's\_theorem (2011).
    
    \bibitem{Hartmann} S. Hartmann and P. Suppes,  \grq Entanglement, Upper Probabilities and Decoherence in Quantum Mechanics', in: Suaraz et. al. (eds.), {\it EPSA Philosophical Issues in the Sciences: Launch of the European Philosophy of Science Association}, Vol. 2. Berlin: Springer, 93--103, (2010). 
    
    \bibitem{Jarrett}  J. P. Jarrett,  \grq On the physical significance of the locality conditions in the Bell arguments'. \textit{No\^us} 18, 569, (1984).
   
  
  \bibitem{Lewis} P.J. Lewis, \grq Retrocausal quantum mechanics: Maudlin's challenge revisited'. \textit{Studies in History and Philosophy of Modern Physics} 44, 442--449, (2013). 
    
    \bibitem{Maudlin} T. Maudlin, \textit{Quantum Non-Locality and Relativity}, Third Ed., Malden, Oxford: Wiley-Blackwell (2011).

\bibitem{Norsen} T. Norsen, \grq Local causality and completeness: Bell vs. Jarrett.' \textit{Foundations of Physics} 39.3, 273--294, (2009).

    \bibitem{Price}H. Price, {\it Time's Arrow \& Archimedes' Point. New Directions for the Physics of Time}. Oxford University Press, New York, Oxford (1996). 
   
   \bibitem{CHSH} A.  Shimony, M.A. Horne and J.F. Clauser,  \grq Proposed Experiment to Test Local Hidden-Variable Theories', \textit{Phys. Rev. Lett.} 23, 880--884, (1969).
   
   \bibitem{CHSH2} A.  Shimony, M.A. Horne and J.F. Clauser, \grq Comment on `The Theory of Local Beables', in: Bonsack,F.(ed) \textit{ Epistemological Letters Hidden Variables and quantum Uncertainty} No. 9 (1976); reprinted in \grq An Exchange on Local Beables', \textit{Dialectica}, Vol. 39, 97--101 (1985).
   
   \bibitem{Salart}{D. Salart, A. Baas, C. Branciard, N. Gisin and H. Zbinden, \grq Testing the speed of `spooky action at a distance'. \textit{Nature} 454, 861--864, (2008).}
   
 \bibitem{Stapp}{H.P. Stapp, \grq Bell's theorem and world process'.  \textit{Il Nuovo Cimento} 40B(29), 270--276, (1975).} 

\bibitem{Sutherland} R. I. Sutherland,  \grq Bell's theorem and backwards-in-time causality'. \textit{International Journal of Theoretical Physics} 22, 377--384, (1983).

\bibitem{Sutherland2}R. I. Sutherland, \grq Causally symmetric Bohm model'. \textit{Studies in History and Philosophy of Science Part B: Studies in History and Philosophy of Modern Physics} 39.4, 782--805, (2008).

  \bibitem{Tumulka} R. Tumulka,  \grq A relativistic version of the Ghirardi-Rimini-Weber model'. \textit{Journal of Statistical Physics} 125, 825--844, (2006).
 
    \bibitem{WF45}{J. A. Wheeler and R. P. Feynman. \grq Interaction with the absorber as the mechanism of
radiation'. \textit{Rev. Mod. Phys.}, 17(2-3), 157, (1945).}

\bibitem{WF49}{J. A. Wheeler and R. P. Feynman.\grq  Classical electrodynamics in terms of direct inter-particle action'. \textit{Rev. Mod. Phys.}, 21(3), 425, (1949).}



\end{thebibliography}
\end{document}